# The doping effect of multiwall carbon nanotube on $MgB_2$/Fe superconductor wire

**J. H. Kim[a)], W. K. Yeoh, M. J. Qin, X. Xu, and S. X. Dou**

*Institute for Superconducting and Electronic Materials, University of Wollongong, Wollongong, Northfields Avenue, NSW 2522, Australia*

**ABSTRACT**

We evaluated the doping effect of two types of multiwall carbon nanotube (CNT) with different aspect ratios on $MgB_2$/Fe monofilament wires. Relationships between microstructure, magnetic critical current density ($J_c$), critical temperature ($T_c$), upper critical field ($H_{c2}$), and irreversibility field ($H_{irr}$) for pure and CNT doped wires were systematically studied for sintering temperature from 650°C to 1000°C. As the sintering temperature increased, $T_c$ for short CNT doped sample slightly decreased, while $T_c$ for long CNT doped sample increased. This indicates better reactivity between $MgB_2$ and short CNT due to its small aspect ratio and substitution of carbon (C) from short CNT for boron (B) occurs. In addition, short CNT doped samples sintered at the high temperatures of 900°C and 1000°C exhibited excellent $J_c$, and that value was approximately $10^4$ A/cm$^2$ in fields up to 8 T at 5 K. This suggests that short CNT is a promising carbon source for $MgB_2$ superconductor with excellent $J_c$. In particular, inclusion of nanosized MgO particles and substitution of C into the $MgB_2$ lattice could result in strong flux pinning centers.

[a)]Electronic mail: jhk@uow.edu.au

## INTRODUCTION

MgB$_2$ is one of the most promising materials for superconductor applications.[1] The high transition temperature of 39 K gives us many advantages for industrial applications where cooling costs and system stability are of concern. The low material cost of magnesium (Mg) and boron (B) is an additional advantage. Furthermore, thanks to its large coherence length and two-gap superconductivity, the opportunity exists to enhance critical current density ($J_c$) and upper critical field ($H_{c2}$) by doping with a wide variety of additives or with substitutional dopants.[2,3] A number of carbon sources, such as SiC, C, B$_4$C, and carbon nanotube (CNT), have shown positive effect on $J_c$ and $H_{c2}$, but result in a large drop in critical temperature ($T_c$).[4-9] Among the various carbon sources, the CNT are particularly interesting as their special geometry may induce effective superconductivity.

It is widely known that the CNT have good electrical, thermal, and mechanical properties, although they exhibit some particularities because of their high aspect ratio and nanometer scale diameters.[10] Many properties of CNT are directly influenced by the way of the graphene sheets are wrapped around. The most important CNT structures are those of single walled nanotubes (SWNTs) and multiwalled nanotubes (MWNTs). A SWNT is considered as a cylinder with only one wrapped graphite sheet. MWNT are similar to a collection of concentric SWNTs. The length and diameter of these structures differ a lot from those of SWNTs and their properties are also very different. Specifically, MWNTs can carry current densities up to $10^9$ - $10^{10}$ A/cm$^2$ and have good thermal conductivity of 3000 W/mK.[11,12] These properties could benefit heat dissipation and thermal stability for MgB$_2$ wires. In addition, MWNT can be used for both the carbon source and as mechanical supports because of its high axial strength.[13] Therefore, including MWNT in MgB$_2$ wire is attractive for superconducting applications.

In the literature, Wei *et al.* have studied the superconductivity of MgB$_2$-CNT composites.[14] Dou *et al.* studied the doping effects of CNT on the $T_c$, $J_c$, and lattice parameters of MgB$_{2-x}$C$_x$ with x=0, 0.05, 0.1, 0.2, and 0.3.[7] The substitution of C from CNT was found to enhance $J_c$ in magnetic field but depressed $T_c$. Recently, Yeoh *et al.* reported that the effect of CNT doping on $H_{c2}$ and irreversibility field ($H_{irr}$).[15] However, the effects of CNT shape (aspect ratio: length over diameter) on the superconductivity of MgB$_2$ were not systematically evaluated. In addition, these researchers estimated the fundamental properties from pellet samples. In this study, we fabricated MgB$_2$/Fe monofilament wires doped with different length CNTs and evaluated the $T_c$, magnetic $J_c$, microstructures, $H_{c2}$, and $H_{irr}$. The resulting properties were compared with a pure MgB$_2$/Fe sample.

**EXPERIMENTAL PROCEDURE**

MgB$_2$/Fe monofilament wires were prepared by an *in-situ* reaction process and powder-in-tube method. Powders of Mg (99%), amorphous B (99%), and CNT (>95%) were used as starting materials. We prepared two kinds of multi-walled CNT as carbon sources for comparison. One had a length (L) of 0.5-200 μm and outer diameter (O.D.) of <8 nm, and the other had with L of 0.5 μm and O.D. of 20-30 nm (hereafter called long and short CNT, respectively, referring to the length of the CNT). These powders were well mixed with Mg and B powders, having the composition of MgB$_{1.8}$C$_{0.2}$. The mixed powders were packed into iron (Fe) tubes with a L of 140 mm. The Fe tube had an O.D. of 10 mm and an inner diameter (I.D.) of 8 mm. The packing process was carried out in air. Both ends of the tubes were sealed with aluminium pieces. The composites were drawn to O.D. of 1.42 mm. These fabricated wires were sintered at 650-1000°C for 30 min under high purity argon (Ar) gas. The heating rate was 5°C/min. After sintering, the wires were furnace-cooled to room temperature. A pure MgB$_2$/Fe wire was also fabricated for comparison by applying the same processing. The volume fraction of the superconducting core in the final wire was approximately 48%.

The microstructure of the wires was evaluated by scanning electron microscope (SEM), transmission electron microscope (TEM), and x-ray diffraction (XRD). Magnetization and ac susceptibility were measured by a physical property measurement system (PPMS, Quantum Design). The magnetic critical current density ($J_c$) was derived from the height of the magnetization loop using Bean's model. Magnetic fields up to 8.0 T were applied parallel to the wire axis of the MgB$_2$ core at 5 K and 20 K. To calculate the magnetic $J_c$ of an MgB$_2$/Fe wire, we first removed the Fe sheath by a mechanical process because this material has ferromagnetic properties. All samples were made to the same size of an O.D. of 0.98 mm and L of 2.5 mm, in order to reduce the sample size effect.[16] Critical temperature ($T_c$) was defined as the onset temperature at which diamagnetic properties were observed. The resistance ($R$) versus temperature ($T$) curves were evaluated in magnetic field up to 8.7 T by the standard four-probe method using PPMS. From these curves, we defined the $H_{c2}$ and $H_{irr}$ as $H_{c2}=0.9R(T_c)$ and $H_{irr}=0.1R(T_c)$, respectively.

**RESULTS AND DISSCUSSION**

Figure 1 shows the XRD patterns of short and long CNT doped $MgB_2$/Fe wires with nominal composition of $MgB_{1.8}C_{0.2}$ after sintering at various temperatures for 30 min. XRD measurements were performed on the ground $MgB_2$ core samples. This figure also shows the XRD patterns of both short and long CNT powders for comparison. As can be seen in this figure, the diffraction patterns of both CNTs showed a broad peak, indicating an amorphous-like phase. Due to their intrinsic nature, the features of XRD patterns of CNTs were close to those of graphite.[17] These phases were not clearly observed in the XRD patterns of CNT doped $MgB_2$/Fe after sintering. However, small amount of un-reacted CNT could exist in the $MgB_2$ core, because the CNT peak overlaps the $MgB_2$ (101) peak at $2\theta \approx 42.5$. In addition, a background-like feature indicates amorphous phase at low $2\theta$ for CNT doped $MgB_2$ wire, suggesting that un-reacted CNT may remain within the $MgB_2$ core. It has been reported that full substitution of C for B site could not be completed, even at a sintering temperature of $1000^oC$.[18] We observed that all samples sintered at different temperatures from $650^oC$ to $900^oC$ seem to be well developed $MgB_2$ with small amounts of MgO. It can be observed that the XRD patterns of the CNT doped $MgB_2$/Fe wire are almost independent of the sintering temperature. For the $1000^oC$ sintering temperature, however, we observed diffraction peaks of Mg and $Fe_xB$ ($Fe_2B$, FeB) for $MgB_2$/Fe wire.[19,20] $Fe_2B$(FeB) may be formed due to chemical reactions between the Fe tube and the $MgB_2$, which results in a small amount of remnant Mg phase. Another possibility is that the loss of Mg due to the high sintering temperature, results in excess B, which reacts with Fe tube to form $Fe_2B$(FeB). From SEM observation, we could see a thin layer in the contact area between the Fe tube and the Mg. Thus, it is difficult to maintain the stoichometry of $MgB_2$ at high sintering temperature.

From the XRD patterns, we also evaluated the full width at half maximum (FWHM) of the (100) peak ($2\theta \approx 33.5^o$) for pure and CNT doped $MgB_2$/Fe wires. The FWHM is related to the crystallinity and lattice distortion. The FWHM of pure wire decreased as the sintering temperature increased. The FWHM values for pure wire were calculated to be $0.366^o$, $0.320^o$, $0.326^o$, $0.308^o$, and $0.250^o$ at sintering temperatures of $650^oC$, $700^oC$, $800^oC$, $900^oC$, and $1000^oC$, respectively. This result can be explained by the improved of crystallinity of the $MgB_2$ core, indicating that the grain size increases with sintering temperature.[21] The FWHM behaviour of long CNT doped wire as a function of sintering temperature was similar to that of pure wire. However, FWHM of short CNT doped wire increased with annealing temperature. The FWHM was $0.350^o$, $0.392^o$, and $0.400^o$ at annealing temperature of $650^oC$, $700^oC$, and $900^oC$, respectively. This indicates that there are lattice distortion due to short CNT doping occurs.

Figure 2 shows DTA traces of pure and CNT doped $MgB_2$/Fe wires fabricated by an *in-situ* process. In order to maintain the same sintering conditions, heating rates were 5°C/min under high purity Ar gas. For pure $MgB_2$/Fe wire, i.e., Mg+2B/Fe, the first exothermal peak was observed at around 457°C, which corresponds to the formation of MgO and melting of $B_2O_3$.[22] The melting temperature of $B_2O_3$ is approximately 450°C. The MgO formation is considered to be due to reaction between melting $B_2O_3$ and Mg, i.e., Mg reacts with $B_2O_3$ to form MgO+B. The $B_2O_3$ as second phase is related to purity of starting powders. The other reason of the existence of MgO in the final wires is because the surface of the commercial Mg powder has partially oxidized. Thus, starting powders with high purity are important for the superconducting properties. The second exothermal peak at 619°C was due to the reaction between melting Mg and amorphous B to form $MgB_2$. These results are consistent with those of Goldacker *et al.*[22] For CNT doped wires, however, the first exothermal peaks were slightly shifted to the higher temperature direction. This is related to the additional impurity due to CNT doping in the starting powders. It is to be noted that the reaction temperature of $MgB_{1.8}C_{0.2}$ formation is almost independent of the aspect ratio between long and short CNTs.

A narrow endothermic peak around 910°C is related to a structural transition of Fe sheath material. According to the Fe-$Fe_3C$ phase diagram, α-Fe (ferrite) with BCC structure transforms into γ-Fe (austenite) with FCC structure at around 912°C.[23] In general, the tetrahedral interstitial site for ferrite is relatively bigger than the octahedral site. Atoms with a maximum radius of 0.35 Å can fit. On the other hand, the octahedral site for austenite is relatively bigger than the tetrahedral site. Atoms with a maximum radius 0.51 Å can now fit. For this reason, it is easy for Fe and B to react above the annealing temperature of 912°C. Iron Boride ($Fe_2B$, $2\theta \approx 45.1°$), especially, can be easily formed at a Fe/Mg ratio greater than 0.02.

The critical temperature ($T_c$) for pure, short, and long CNT doped $MgB_2$/Fe wires are presented in Figure 3. For pure $MgB_2$/Fe wire, $T_c$ increased systematically as the sintering temperature increased. This is related to the improvement of crystallinity of the $MgB_2$ core. It is to be noted that the $T_c$ variations of the doped wires are significantly different for the two types of CNTs. As the sintering temperature increased, $T_c$ for the short CNT sample slightly decreased, on the other hand, $T_c$ for the long CNT sample increased. In addition, $T_c$ of short CNT doped wire was slightly lower than for the long one at the same sintering temperature. For example, $T_c$ values of long CNT doped wire were 37.2 K, 37.0 K, and 38.1 K at 650°C, 800°C, and 1000°C, respectively, while the corresponding values for the short one were 36.2 K, 36.5 K, and 35.7 K. This result is considered to be due to better reactivity between $MgB_2$ and short CNT. Since the short CNT has a smaller aspect ratio. On the other hand, long CNT has a higher aspect ratio and tends to entangle, preventing homogeneous mixing in $MgB_2$. In order to explore other possibilities, we also estimated the core density of pure and CNT doped wires. The core density improved in all samples with sintering temperature. The mean values were calculated to be 1.33 g/cm$^3$, 1.42 g/cm$^3$, and 1.42 g/cm$^3$ in pure, short, and long CNT doped wire, respectively. Even though the density of our pure sample (1.33 g/cm$^3$) was not closed to the theoretical density (2.62 g/cm$^3$), the density values of both CNT doped wires were the same. We found that the actual doping reactivity between CNT and MgB$_2$ is strongly influenced by the shape of carbon source.

Figure 4 shows the magnetic dependence of $J_c$ for short and long CNT doped wires at 5 K and 20 K. The best $J_c(B)$ result of pure MgB$_2$ wire is also plotted for comparison and was estimated to be 0.34 MA/cm$^2$ at 20 K in self-field. It was observed that $J_c$ for the short CNT doped sample was much higher than that for the long one. Specifically, the short CNT doped samples sintered at the high temperatures of 900$^o$C and 1000$^o$C exhibited excellent $J_c$, approximately 10$^4$ A/cm$^2$ in fields up to 8 T at 5 K. In addition, $J_c$ of the sample sintered at the highest temperature showed a crossover with that of the low temperature one around 5 T. This result indicates that flux pinning after high temperature sintering was enhanced by a large amount of C substitution. It is to be noted that $J_c$ for the long CNT doped sample was depressed monotonically as the sintering temperature increased. This behaviour is likely to be related to the increasing of MgB$_2$ grain size due to inhomogeneous mixing. As mentioned above, FWHM of the long CNT doped wire decreased with annealing temperature, indicating that grain size increases. Many groups have reported that the increased of grain size in MgB$_2$ is an important factor in the suppression of $J_c$, since grain boundary pinning is important mechanism in controlling $J_c$.[4,24-26] In addition, we observed that there was no crossover around 5 T. This behaviour is similar to that of the pure MgB$_2$ wire, and $J_c$ was <10$^3$ A/cm$^2$ at 8 T and 5 K. Poor field performance for long CNT doped wire is likely to be ultimately related to the poor reactivity and inhomogeneous mixing between MgB$_2$ and long CNT.

In order to evaluate the reactivity, we measured the specific surface area (SSA) of both CNTs and B powder using surface area analyser (NOVA 1000) and the well-known BET (Brunauer-Emmett-Teller) equation. It was observed that the SSA of long CNT was much bigger than that of the short one. The measured SSAs were 31.50 m$^2$/g, 120.43 m$^2$/g, and 408.50 m$^2$/g for B powder, short CNT, and long CNT, respectively. From this result, one would expect long CNT to be more reactivity. However, magnetic $J_c$ of this sample was the same as for the pure one. This is because the surface of CNT is very stable. Even though long CNT has a bigger surface area, compared to short CNT, only the end holes of CNT are effective as reaction site.[27,28] Generally, the properties of the CNTs are influenced by defects sites on the walls and at the ends. The caps at both ends of the CNTs are removed and defects on surface are revealed by purification techniques.

Figure 5 plots the field dependence of the volume pinning force, $F_P = J \times B$, of short and long CNT doped sample at 20 K. The $F_p$ is normalized by the maximum volume pinning force, $F_{p,max}$, of pure $MgB_2$. When $0 < H < 0.5$ T, the shapes of these plots were almost the same as that of the pure sample. For the wires at $0.5 < H < 2.0$ T, it was observed that the pure $MgB_2$ sample is much more conducive to obtaining a strong $F_P(H)$. At $H > 2.0$ T, however, the $F_P(H)$ of the short CNT doped sample is larger than that of the pure and long CNT doped sample. This result indicates that the $F_P(H)$ of the $MgB_2$ tapes was improved by the doping effect of short CNT. It is to be noted that $F_P(H)$ behavior of the long CNT doped sample was similar to that of the short CNT doped sample. A similar flux pinning mechanism is likely to be involved for both CNT doped samples.

Figure 6 shows TEM images for short, (a) and (b), and long, (c) and (d), CNT doped $MgB_2$ sintered at 900°C. In all the specimens, the powders were suspended on 'lacey carbon grids' - that is a network of carbon filaments. As can be seen from the figures, the short CNT doped samples consisted of crystalline $MgB_2$ grains of 100-200 nm, small $MgB_2$ grains of 10-20 nm in size, and nanoparticles. From energy dispersive x-ray (EDX) spectra, these particles were rich in oxygen. However, it is difficult to know the exact composition of the nanoparticles. These nanoparticles could be related to the CNT doping. Some areas (such as 6(b)) showed CNT sticking out of $MgB_2$ grain and little lumps of CNT. For the long CNT doped sample, on the other hand, we observed large lumps of un-reacted CNT near $MgB_2$ grains (see 6(d)). Because of the entanglement of long CNT due to the high aspect ratio, long CNT is difficult to align along the direction of the wire axis during mechanical processing. However, there also appears to be a large number of very small grains and nanoparticles. Diffraction patterns, shown in the top right corners (6(a) and (c)), gave ring type patterns consistent with $MgB_2$. There are extra rings in these patterns, which suggest another phase, i.e., these rings appear to be consistent with MgO. According to two-gap superconductivity, the nanoparticles could enhance the flux pinning effect. Since the *a-b* plane coherence length of $MgB_2$ is approximately 6-7 nm, partial inclusion of nanosized particles could result in strong flux pinning centers.[29]

It is to be noted that there are no nanoparticles within the pure $MgB_2$ matrix. However, the magnetic $J_c$ of pure $MgB_2$ sintered at 700°C was the same as that of the long CNT doped one sintered at 900°C, as mentioned above. This is because a small $MgB_2$ grain size allows the extra grain boundaries to also act as strong flux pinning centers. The pure samples sintered at 700°C consisted of $MgB_2$ grains of 50~150 nm. Yamada *et al.* reported that the small $MgB_2$ grain size was effective in enhancing flux pinning because the grain boundaries of $MgB_2$ represented effective pinning centers, as in the case of A15 metallic superconductors.[25] Fischer *et al.* fabricated $MgB_2$ wires and bulks by a mechanical alloying method.[30] They obtained one of the highest $J_c$ values without any other element materials added. The very fine-grained nanocrystalline microstructure of the superconducting phase seems to be responsible for these excellent $J_c$ values.[31]

The temperature dependence of $H_{c2}$ and $H_{irr}$ for short and long CNT doped MgB$_2$ sintered at 900°C is shown in Figure 7. As can be seen from the figure, the $H_{c2}$ of the short CNT doped sample was higher than that of the long one, which was similar to that of the pure MgB$_2$. The short CNT doped sample demonstrated larger $dH_{c2}/dT$, compared with the pure and the long one. These are believed to increase the intra-band scattering, shorten the mean free path and coherence length.[6] In addition, $H_{c2}$-$T$ curves of all samples show similar behaviour to those of the $H_{irr}$-$T$ curves. We also evaluated the resistivity of pure and CNT doped samples. The values were calculated to be 38.5 μΩ·cm, 57.5 μΩ·cm, and 2.60 μΩ·cm at 40 K for the pure, short, and long CNT doped wire, respectively. The short CNT doped samples showed a relatively higher value of resistivity at 40 K. The increased resistivity for the short CNT doped sample could be due to the greater C substitution or to higher sample density. However, the long CNT doped sample had a lower resistivity, compared to the pure sample. From the resistivity curve, the residual resistivity ratio (*RRR*) was estimated to be 2.02, 1.98, and 1.63 for the pure, long, and short CNT doped samples, respectively.[32]

**CONCLUSION**

In this study, we evaluated the doping effect of CNT with different aspect ratios on MgB$_2$/Fe monofilament wires. Relationships between microstructure, $J_c$, $T_c$, $H_{c2}$, and $H_{irr}$ for both short and long CNT doped wires were systematically studied. We observed that all samples sintered at different temperature from 650 to 900°C seem to be a well-developed MgB$_2$ with small amount of MgO. However, the FWHM of the (100) peak of short CNT doped wire increased with sintering temperature. This indicates that lattice distortion and depressed crystallinity have occurred due to short CNT doping. Specifically, short CNT doped samples sintered at high temperatures exhibited excellent $J_c$, ~$10^4$ A/cm$^2$ up to 8 T at 5 K. The $J_c$ of the sample sintered at the highest temperature showed a crossover with that of the low temperature one around 5 T. This result indicates that flux pinning for samples produced at a high sintering temperature was enhanced by short CNT doping. In addition, the short CNT doped sample presented a larger $dH_{c2}/dT$, compared with the pure and the long one. The short CNTs are thus believed to increase the intra-band scattering, shorten the mean free path and coherence length. The short CNT is a promising carbon source for MgB$_2$ superconductor with excellent $J_c$.

**ACKNOWLEDGMENTS**

The authors thank Dr. T. Silver, Dr. J. Horvat, Dr. R. H. T. Wilke, and Mr. R. Kinnell for their helpful discussions. This work was supported by the Australian Research Council, Hyper Tech Research Inc., USA, Alphatech International Ltd., NZ, and the University of Wollongong.

**FIGURE CAPTIONS**

**Figure 1.** The X-ray diffraction patterns for short and long CNT doped $MgB_2$ wires with nominal composition of $MgB_{1.8}C_{0.2}$ after sintering at various temperatures for 30 min. The patterns of the precursor CNT powders are also shown for comparison.

**Figure 2.** The DTA curves for pure and CNT doped wires processed with a heating rate of 5°C/min.

**Figure 3.** The relationships between $T_c$ and sintering temperature for pure and CNT doped wires.

**Figure 4.** Critical current densities for (a) short CNT doped and (b) long CNT doped wire sintered at various temperatures for 30 min. Magnetic $J_c$ curves of pure $MgB_2$ wires sintered at 700°C for 30min are also shown for comparison.

**Figure 5.** The field dependence of the volume pinning force, $F_P = J \times B$, of short and long CNT doped $MgB_2$ wires at 20 K. The $F_p$ is normalized by the maximum volume pinning force, $F_{p,max}$, of pure $MgB_2$.

**Figure 6.** TEM images for short CNT doped (a) and (b), and long CNT doped $MgB_2$ (c) and (d) wires with the nominal composition of $MgB_{1.8}C_{0.2}$. The insets in (a) and (c) show ring-type diffraction patterns consistent with $MgB_2$ + another phase, possibly MgO.

**Figure 7.** Temperature dependence of $H_{c2}$ and $H_{irr}$ for short and long CNT doped $MgB_2$ sintered at 900°C. The $H_{c2}$ and $H_{irr}$ of pure $MgB_2$ wires sintered at 700°C for 30min are also shown for comparison.

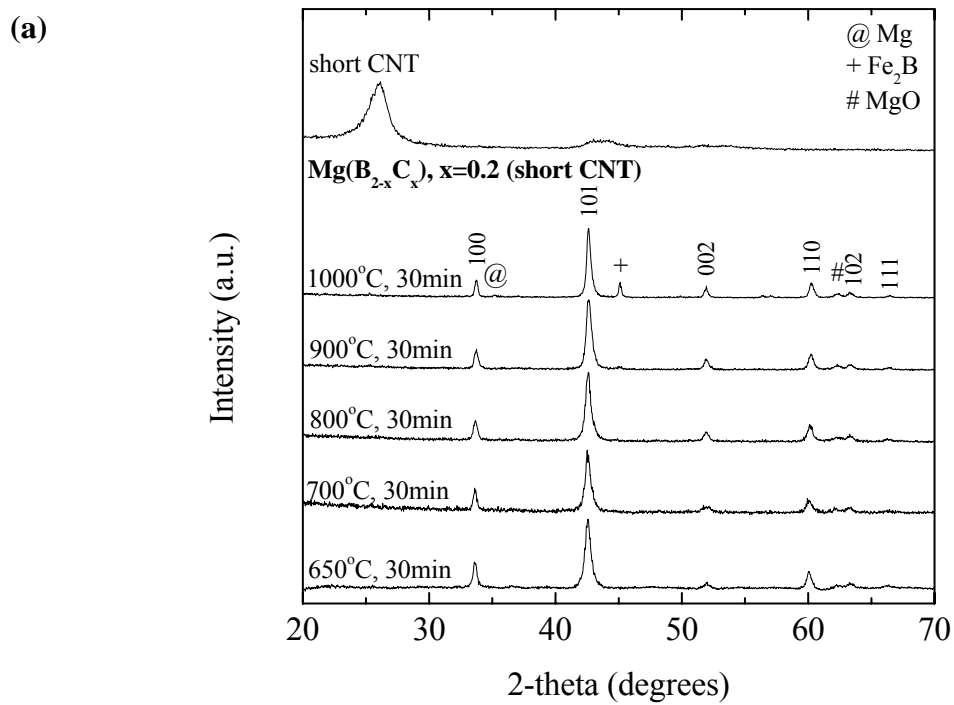

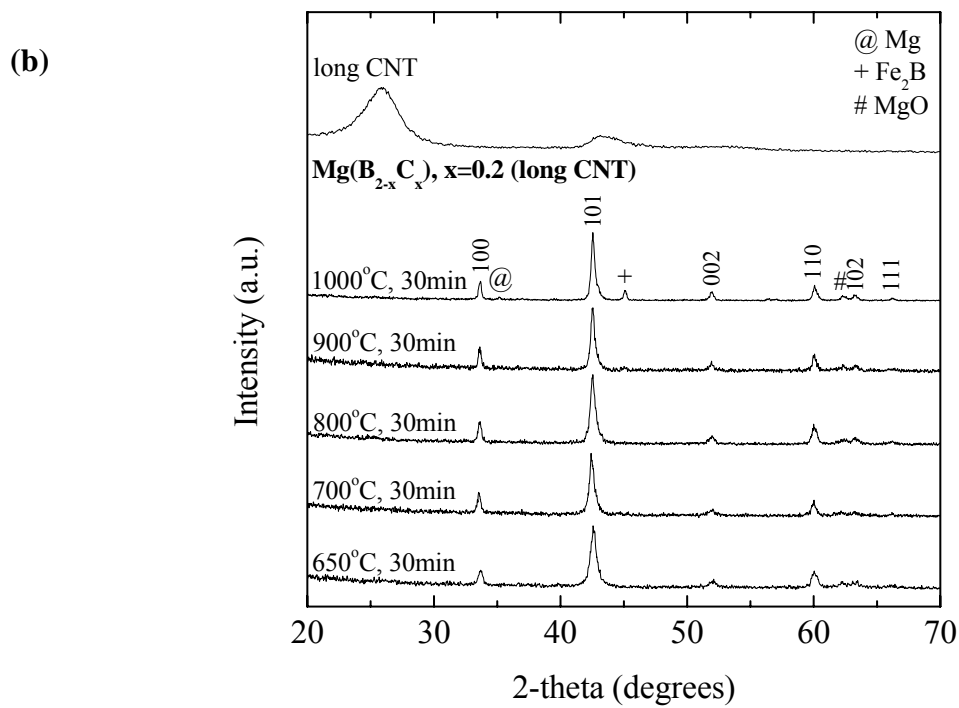

**Figure 1.**

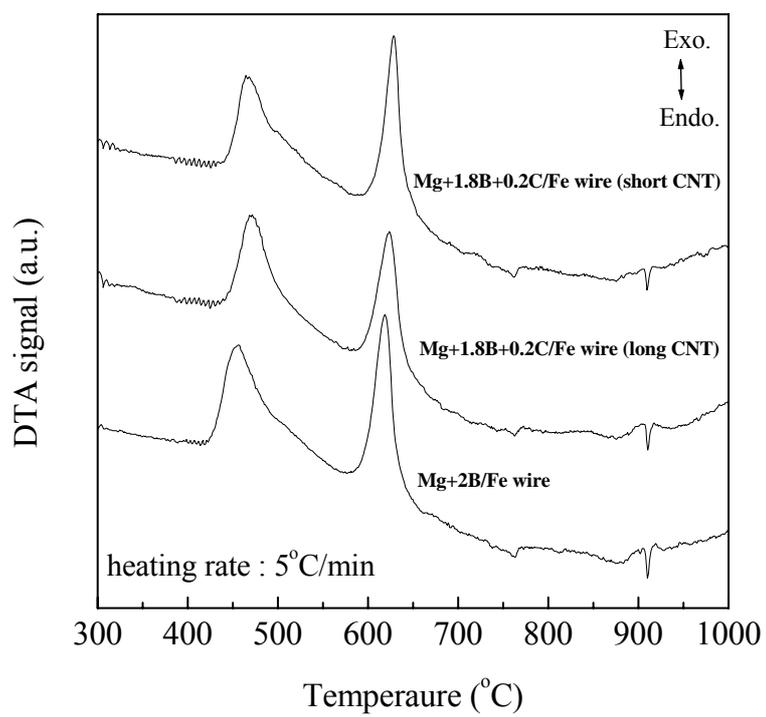

**Figure 2.**

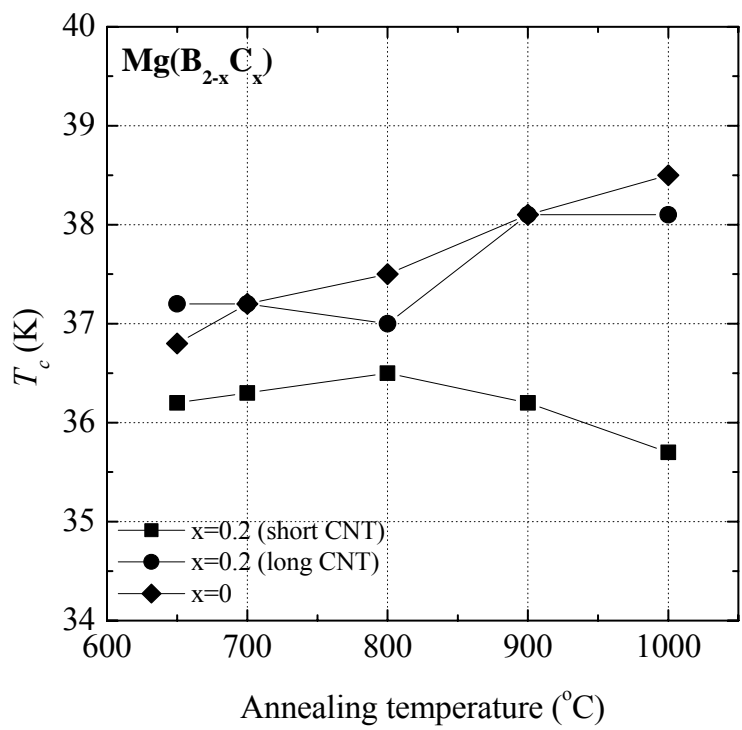

**Figure 3.**

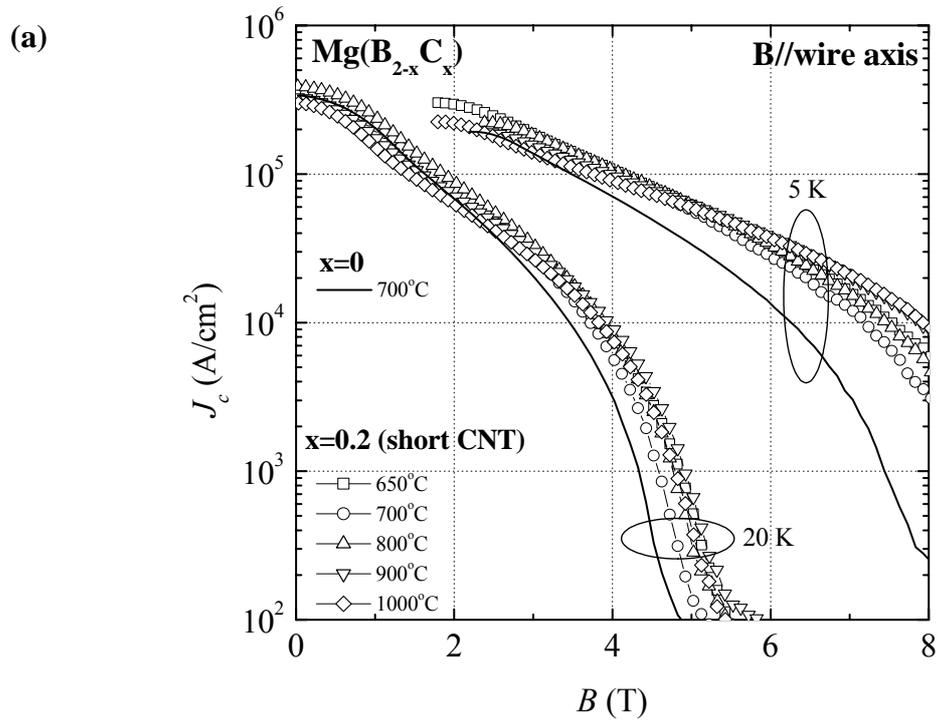

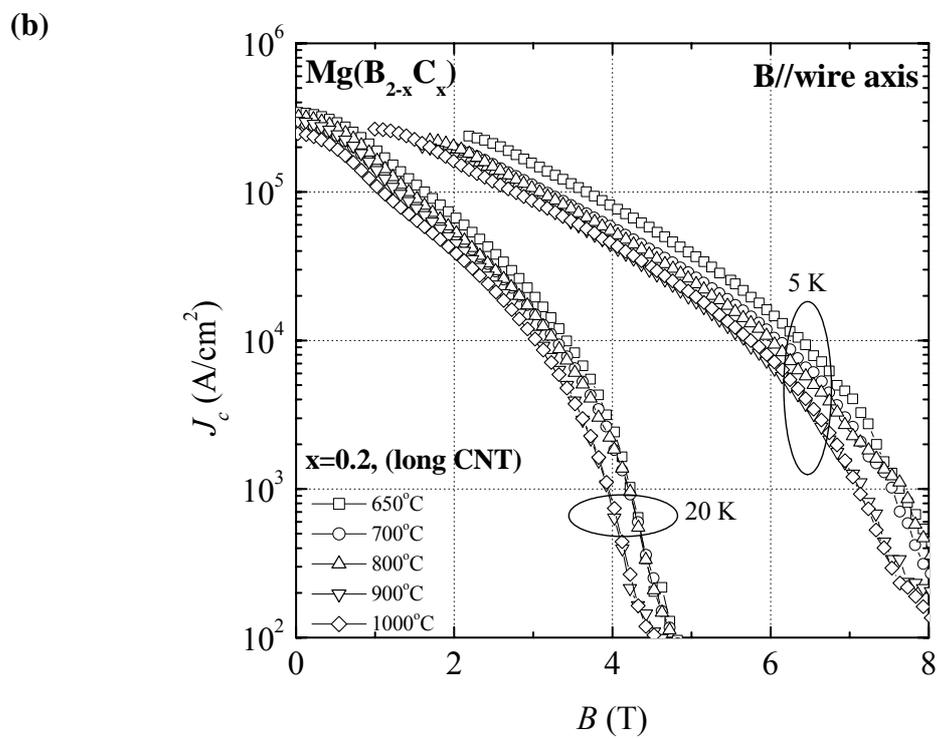

**Figure 4.**

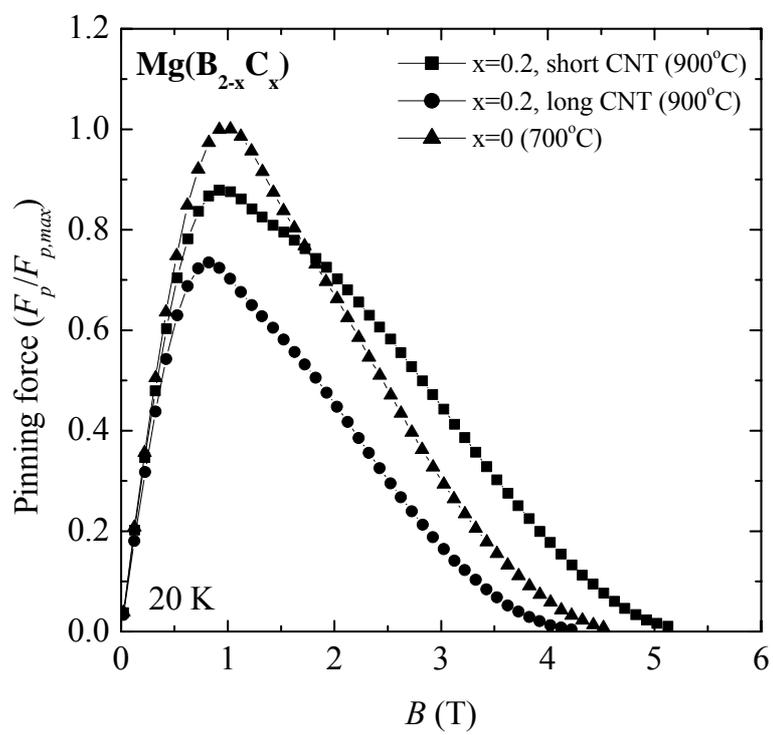

**Figure 5.**

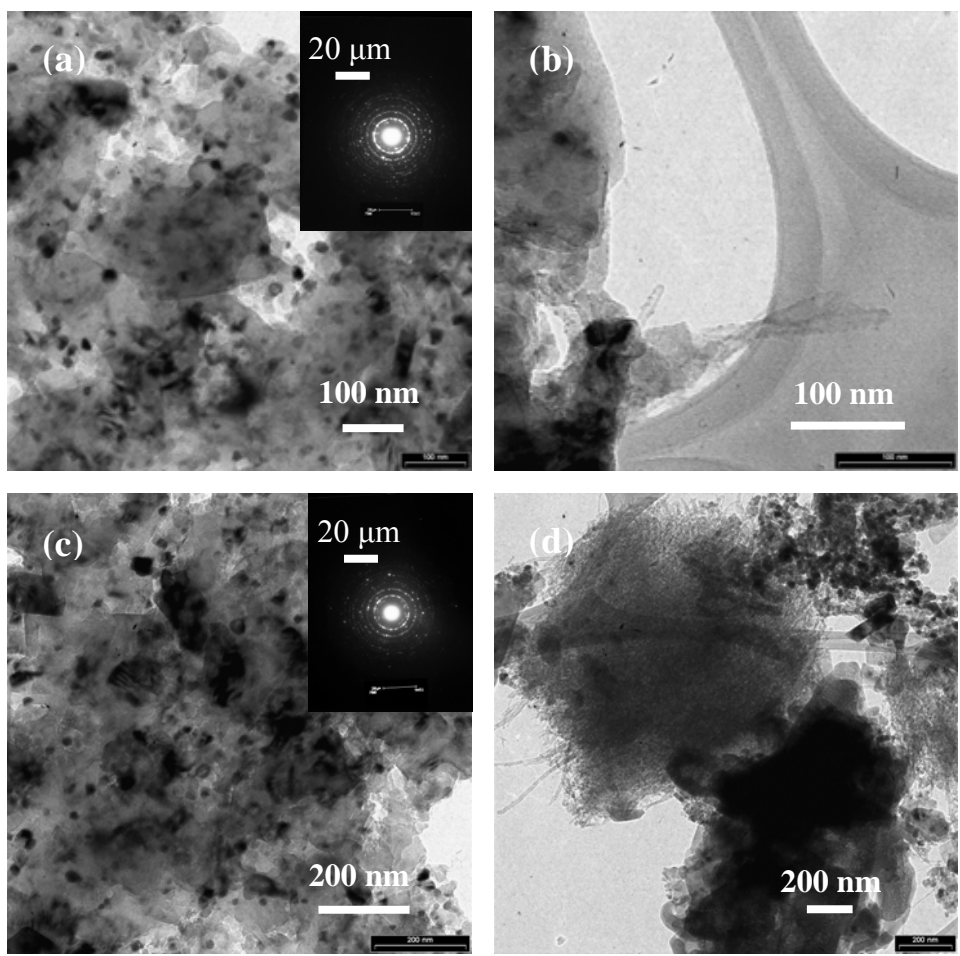

**Figure 6.**

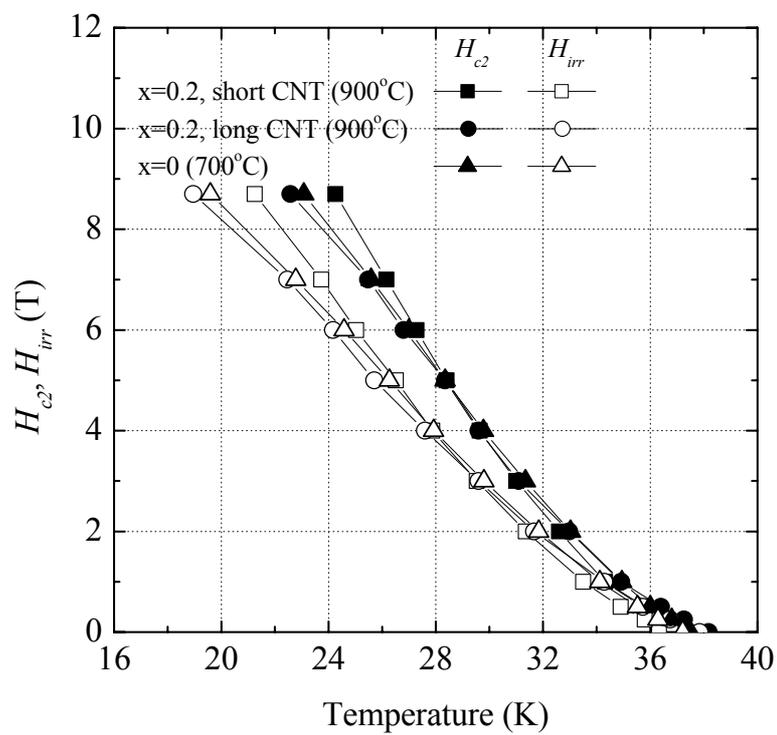

**Figure 7.**